\documentclass[aps,prx,10pt,amsmath,amssymb,floatfix,superscriptaddress,notitlepage
]{revtex4}

\usepackage{dcolumn}
\usepackage{graphicx}
\newcolumntype{d}[1]{D{.}{.}{#1}}
\usepackage{hyperref}
\begin{document}
% \begin{frontmatter} % The preamble begins here.

%
%\pretitle{Pretitle}
\title{Small Angle Neutron Scattering in McStas: optimization for high throughput virtual experiments}
% \title{Small Angle Neutron Scattering form factor models for sample description in McStas}

% \runtitle{SANS in McStas: optimization for high throughput virtual experimentss}
%\subtitle{Subtitle}

% \begin{aug}
\author{Jose Robledo}
\email{j.robledo@fz-juelich.de}%
\affiliation{Jülich Centre for Neutron Science 2, Forschungszentrum Jülich, Wilhelm-Johnen-Straße, 52428, Jülich, Germany}
\author{Klaus Lieutenant}
\email{k.lieutenant@fz-juelich.de}
\affiliation{Jülich Centre for Neutron Science 2, Forschungszentrum Jülich, Wilhelm-Johnen-Straße, 52428, Jülich, Germany}
\author{Peter Willendrup}
\email{pkwi@fysik.dtu.dk}
\affiliation{Data Management and Software Centre (DMSC), European Spallation Source,
2800 Kongens Lyngby, Denmark}
\affiliation{Physics Department, Technical University of Denmark,
2800 Kongens Lyngby, Denmark}
% \end{aug}

\begin{abstract}
In this work we present the development of small angle scattering components in McStas that describe the neutron interaction with 70 different form and structure factors. We describe the considerations taken into account for the generation of these components, such as the incorporation of polydispersity and orientational distribution effects in the Monte Carlo simulation. These models can be parallelized by means of multi-core simulations and graphical processing units (GPUs). The acceleration schemes for the aforementioned models are benchmarked, and the resulting performance is presented. This allows for the estimation of computation times in high-throughput virtual experiments. The presented work enables the generation of large datasets of virtual experiments that can be explored and used by machine learning algorithms.  
\end{abstract}

\maketitle
% \begin{keyword}
% \kwd{virtual experiments}
% \kwd{small angle neutron scattering}
% \kwd{acceleration}
% \kwd{graphical processing unit}
% \kwd{machine learning}
% \end{keyword}
% \end{frontmatter}

\section{Introduction}

McStas \cite{mcstas, mcstas2} is an open-source package for Monte Carlo (MC) neutron tracing simulations widely used by the neutron scattering community for instrument design in neutron facilities and also for educational purposes. A recent trend in the use of MC simulation software, inspired by increasing computing power, is the generation of large virtual experimental datasets. These datasets can then be explored using machine learning (ML) algorithms to learn underlying structures that may provide insight into novel data analysis techniques and instrument design optimization \cite{robledo2024}. If the system is anisotropic, then the expression for the Small Angle Scattering (SAS) model now depends on the orientation of the particle relative to the incident beam, and therefore a shape-specific modulation of the scattering pattern as a function of the scattering vector $\vec Q$ is present and should be adequately described \cite{anisosas}.  

Large-scale datasets can be constructed by varying simulation parameters in virtual experiments of neutron instruments. The information that can be extracted from such datasets depends greatly on the possibility that the MC software grants in exploring the simulation's hyper-parameter space. 
This space consists mainly of parameters related to the instrument configuration in a given MC simulation, as well as on parameters that describe the neutron interaction with the sample.
In McStas, the complexity of the simulation's parameter space is determined by the available components that describe the  instrument and sample in a virtual experiment. This work aims to expand the latter by introducing 70 new SAS model components that can be used in McStas to describe a sample in a virtual neutron experiment. 

In a first approximation and for simple problems where only first-order interactions and no secondary productions are taken into account \cite{MCparallel?} Monte Carlo neutron tracing simulations are embarrassingly parallel \cite{embarrassingparallel}. That is, each neutron generated from the source distribution by means of MC choices is independent of the previously generated neutrons, and therefore the task of generating neutrons can be parallelized. The propagation of the neutrons in the beamline and its interaction with instrument components can also be parallelized because they are also independent. If only single scattering events are considered, the inclusion of a form factor or structure factor model for the sample description does not change this characteristic. Therefore, large dataset generation can have a significant speed-up if parallelization strategies are used in the MC neutron virtual experiments that generate the data. 

This work describes the inclusion of 70 SAS models in McStas that are optimized for parallel execution. A benchmark of the computation time of these models is also presented to emphasize and give estimates of the computing times necessary to build up big datasets. 

\section{Methods}
\subsection{Form factor and structure factor models}

In scattering theory, the form factor $P(\vec Q)$ of a given model describes the dependence of the intensity of the scattering that is due solely to the shape of the scattering object. In Small Angle Scattering (SAS), the information due to the sample composition is encoded in the excess of scattering length density (SLD), $\Delta \rho$, and the modulation of the intensity due to the relative positions of scattering objects is described by the structure factor $S(\vec Q)$.

In the most general case of scattering from a scattering object of the model $m$, if we define the scattering amplitude as \cite{sasbook}

\begin{equation}
    A_m(\vec{Q}) = \int_{V_m}  \Delta \rho_m(\vec{r}) \exp{\left[-i\vec{Q}\cdot \vec{r}\right]}dV,
\end{equation}
with $V_m$ the volume of the scattering particle, then the measured intensity as a function of the scattering vector $\vec Q$  can be written as 

\begin{equation}
    I_m(\vec{Q}) = \frac{\Phi}{V_m} A_m(\vec{Q})A^*_m(\vec{Q}),
\end{equation}
where $\Phi$ is a scale factor.
In a definition of the form factor $P(\vec{Q})$ as the contribution due solely to the shape of the scattering object, since $\Delta \rho$ depends on the composition and not on the shape of the scatterers, it becomes necessary to separate it from the modulus square of the scattering amplitude $A^*(\vec{Q})A(\vec{Q})$. In most cases this is straight-forward, like in an elastic, isotropic scattering experiment of a dilute mono-disperse system, where the expression reduces to

\begin{equation}
    I_m(Q) = n \Delta \rho^2 V_m^2 P_m(Q) S(Q)
\end{equation}
with $n$ the number density of particles. Since the intensity is related to the modulus squared of the amplitude, then the quantity $\Delta \rho^2$ appears often in the literature, and is referred to as the contrast.
We can define the Small Angle Scattering (SAS) model of a sample including the $\Delta \rho$ dependency and also normalizing it by the volume of the particle so as to easily compare the scattering intensity measured in experiments, i.e.

\begin{equation}
    M_m(\vec{Q}) = \frac{1}{V_m}A_m^*(\vec{Q})A_m(\vec{Q}). \label{eq:form_factor}
\end{equation}
Given that $\Delta \rho$ is present in this expression through $A(\vec{Q})$, the SAS model can be interpreted as describing the coherent scattering amplitudes arising from regions of excess SLD \cite{naturesas}. Finally, each model $m$ considered in this work describes the shape characteristics of the scattering object through the definition of a set of model parameters $\left\{\theta_i | i=1,\dots,n_m\right\}$, therefore the notation $M_m(\vec{Q}|\theta_1,\dots,\theta_{n_m})$ will be used to explicitly refer to $M_m(\vec{Q})$ when the complete model is defined. 

 A total of 70 SAS models have been developed as independent components in McStas \cite{mcstas}, and are available since version 3.4 onward. Tables \ref{table::anisotropic_models} and \ref{table::isotropic_models} present a summary of the models, classified into seven similarity groups labeled A to G. The components have also been imported to the X-ray counterpart McXtrace\cite{mcxtrace} as they describe the small angle scattering characteristics that depend on the sample and not on the incident particle. Each model is described by two functions written in $C$ language: one defining the scattering amplitude $A_m(\vec{Q}|\theta_1,\dots,\theta_{n_m})$ and another the corresponding particle volume $V_m$. Special care is taken in the case of anisotropic models to orient particles relative to the incident beam by means of rotation matrices. The analytical model descriptions were obtained from the SasView \cite{sasview} software package, which is widely used for SAS data analysis. These analytical models are well-known and tested within the SAS community.  Additionally, a Python script has been included in the McStas repository to automatically extract the analytical functions from SasView and create the McStas components for future updates of available SAS models. 

\begin{table}[ht]
\centering
\caption{Anisotropic Small Angle Scattering models included in McStas, inherited from SasView. Models are divided into 7 groups, identified with letters from A to G. For a complete description of each model, visit the Mcstas component documentation and the SasView user documentation (https://www.sasview.org/docs/user/qtgui/Perspectives/Fitting/models/index.html).}
\label{table::anisotropic_models}
\renewcommand{\arraystretch}{1.2}
\setlength{\tabcolsep}{5pt}
\begin{tabular}{>{\bfseries}c>{\raggedright\arraybackslash}p{2.5cm}>{\raggedright\arraybackslash}p{2.5cm}>{\raggedright\arraybackslash}p{2.5cm}>{\raggedright\arraybackslash}p{2.5cm}>{\raggedright\arraybackslash}p{2.5cm}>{\raggedright\arraybackslash}p{2.5cm}}
\hline
\textbf{} & \textbf{A. Cylinders} & \textbf{B. Ellipsoids} & \textbf{D. Paracrystals} & \textbf{E. Parallelepiped} & \textbf{F. Spheres} \\ \hline
1  & Barbell                         & Core shell ellipsoid & Body centered cubic (bcc) & Core shell parallelepiped & Superball \\ 
2  & Capped cylinder                 & Ellipsoid            & Face center cubic (fcc)   & Hollow rectangular prism   &\\ 
3  & Core shell bicelle              & Triaxial ellipsoid   & Simple cubic (sc)         & Parallelepiped             &\\ 
4  & Core shell bicelle elliptical   &                      &                            &   Rectangular Prism                         & \\ 
5  & Core shell bicelle elliptical belt rough &              &                            &                            & \\ 
6  & Core shell cylinder             &                      &                            &                            & \\ 
7  & Cylinder                        &                      &                            &                             &\\ 
8  & Elliptical cylinder             &                      &                            &                            & \\ 
9  & Hollow cylinder                 &                      &                            &                            & \\ 
10 & Stacked disks                   &                      &                            &                            & \\ 
\hline
\end{tabular}
\end{table}

\begin{table}[ht]
\centering
\caption{Isotropic Small Angle Scattering models included in McStas, inherited from SasView. Models are divided into 7 groups, identified with letters from A to G.}
\label{table::isotropic_models}
\renewcommand{\arraystretch}{1.2}
\setlength{\tabcolsep}{5pt}
\begin{tabular}{>{\bfseries}c>{\raggedright\arraybackslash}p{2.5cm}>{\raggedright\arraybackslash}p{2.5cm}>{\raggedright\arraybackslash}p{2.5cm}>{\raggedright\arraybackslash}p{2.5cm}>{\raggedright\arraybackslash}p{2.5cm}}
\hline
\textbf{} & \textbf{A. Cylinders} & \textbf{C. Lamellae} & \textbf{E. Parallelepiped} & \textbf{F. Spheres} & \textbf{G.Shape-Indep.} \\ \hline
1  & Flexible cylinder                                      & Head and tail groups (hg)                   & Hollow rectangular prism with thin walls & Adsorbed layers                & Broad peak              \\  
2  & Flexible cylinder elliptical                           & Caille structure factor (stack caille)      &       & Binary hard spheres            & Correlation length   \\ 
3  & Pearl necklace                                         & hg + Caille structure factor (hg stack caille) &                            & Core shell sphere              & Debye Anderson Brumberger (dab)   \\ 
4  & Pringle                                             & with paracrystal structure factor (stack paracrystal) &                     & Fuzzy sphere                   & Fractal \\
5 & & & &  Linear pearls                  & fractal core shell           \\ 
6 & & &  & Multilayer vesicle             &  Gauss Lorentz Gel\\
7 &&& & Polymer micelle                &     Gaussian peak             \\ 
8 &&& & Raspberry                      & Gel fit                      \\ 
9 &&& & Sphere                         & Guinier                  \\ 
10 &&&& Vesicle                        & Guinier porod             \\ 
11 &&&& Hard sphere                    & Line             \\ 
12&&& & Hayter mean squared approximation & Lorentz       \\ 
13 &&&& Sticky hard sphere             & Mass fractal  \\ 
14 &&&&                                & mass surface fractal                 \\ 
15 &&&&                                & mono gauss coil              \\ 
16 &&&&                                & Peak Lorentz                 \\ 
17 &&&&                                & Polymer excluded volume            \\ 
18&&& &                                & Poly gauss coil         \\ 
19 &&&&                                & Porod            \\
20&&& &                                & Power law         \\
21 &&&&&Spinoidal \\
22 &&&&&square well\\
23 &&&&&Star polymer\\
24 &&&&&Surface fractal \\
25 &&&&&Teubner strey\\
26 &&&&&two lorentzian\\
27 &&&&&Two power law\\
\hline
\end{tabular}
\end{table}

A total of 21 models exhibit anisotropic analytical models, which are indicated in table \ref{table::anisotropic_models}. In an isotropic scattering model, the measured intensity $I$ is proportional to the scattering amplitude $A$, which is solely a function of the modulus of the scattering vector $\vec{Q}$, i.e.
\begin{equation}
    I(\vec{Q}) \propto A_m(\vec{Q}|\theta_1, \dots, \theta_{n_m})=A_m^{iso}(Q|\theta_1, \dots, \theta_{n_m}),
\end{equation}
where $\theta_i$ are the $n_m$ parameters that define the scattering model $m$. This is the case for all 49 models that are in Table \ref{table::isotropic_models}. In contrast, in the anisotropic cases, the measured intensity depends of the relative orientation of the scattering objects with respect to the scattering vector. If the scattering object has no symmetry axis, then the scattering amplitude can be written as
\begin{equation}
    A_m(\vec{Q}|\theta_1, \dots, \theta_{n_m})=A_m^{aniso}(Q_A, Q_B, Q_C|\theta_1, \dots, \theta_{n_m}),
\end{equation}
where $Q_A$, $Q_B$, and $Q_C$ are the projections of $\vec{Q}$ on the $A$, $B$, and $C$ axes of the scattering object. The orientation of the scattering objects can be defined relative to the Monte Carlo simulation global coordinate frame $xyz$ (defining the incoming neutron or X-ray beam) by defining three angles $\theta,\,\phi,\,\Psi$. In this definition, $\theta$ and $\phi$ define the orientation of the $C$-axis of the particle, and the angle $\Psi$ defines the rotation about the $C$-axis. The 12 models from Tables \ref{table::anisotropic_models} and \ref{table::isotropic_models} that belong to this type are: some cylinder models (group A: Core shell bicelle elliptical, Core shell bicelle elliptical belt rough, Core shell parallelepiped),  some ellipsoids (group B: Elliptical cylinder, Triaxial ellipsoid), all paracrystals (group D: BCC paracrystal, FCC paracrystal, SC paracrystal), some parallelepipeds (Group E: Hollow rectangular prism, Parallelepiped, Rectangular prism), and a particular type of sphere model which with a parameter can be transformed into a parallelepiped (Group F: Superball).

If the form of the scattering object has one symmetry axis, suppose $C$, then the scattering amplitude can be calculated by having the projection of the scattering vector on that axis and the projection of the scattering vector onto the plane $AB$, i.e. 
\begin{equation}
    A_m(\vec{Q}|\theta_1, \dots, \theta_{n_m})=A_m^{aniso}(Q_{AB}, Q_C| \theta_1, \dots, \theta_{n_m}).
\end{equation}
There are 9 of such cases in the developed models: most cylinders (Group A: Barbell, Capped cylinder, Core shell bicelle, Core shell cylinder,  Cylinder,  Hollow Cylinder, Stacked disks) and ellipsoids (Group B: Ellipsoid, Core shell ellipsoid).

To test the developed models, we simulated the KWS-1 SANS beamline at FRM-II in McStas. This simulation consisted of a monochromatic source, guides and slits to propagate the neutron beam until the sample with defined divergence, one of the SANS modules listed in tables \ref{table::anisotropic_models} and \ref{table::isotropic_models}, and a 2D Position Sensitive Detector (PSD) at a defined distance from the sample. For an accurate description of the McStas instrument, the simulation description can be found in the Supplementary Material of Ref. \cite{robledo2024}.

\subsection{polydispersity and orientational distributions}\label{sec:poli}

The components presented in this work offer the possibility of defining distributions for some model parameters in which this can be relevant. This includes polydispersity in all distances and radii parameters and orientational distributions in $\theta$, $\phi$, and $\Psi$ orientation angles (described in the previous section) of anisotropic models. These features are included by considering a different size and orientation of the scattering object for each scattering event, which are randomly chosen from defined distributions. 

For the moment, the polydispersity of a parameter $r_a$ is incorporated through the random sampling of its value at each neutron interaction from a Gaussian distribution, i.e.
\begin{equation}
    r_a \sim N(r_a,\sigma_a^2),
\end{equation}
 with $r_a$ and  $\Delta r = \sigma_r/r$ provided by the user. Radii and distances are always checked to be positively defined, else a new iteration of the random choice is performed. 

The orientational distributions are also defined on each neutron interaction by random sampling from uniform distributions of the angles $\theta$, $\phi$ and $\Psi$ as follows,

\[\begin{array}{ccc}
    \theta & \sim & U[0,\pi], \\
    \phi & \sim & U[0,2\pi], \\
    \Psi & \sim & U[0, 2\pi].\\
\end{array}\]
Increasing the number of incident neutrons to the sample in a simulation improves the Monte Carlo estimates of  polydispersity and orientational distribution contributions to the scattering pattern measured at a detector.

\subsection{Acceleration schemes}\label{sec:acc}

The developed components allow for multi-thread parallelization by means of Message Passing Interface (MPI) \cite{mpi41} and can also be deployed to the Graphical Processing Unit (GPU) for higher order acceleration. This is managed internally in each component using OpenACC \cite{openacc}. 

\begin{figure}[h!]
    \centering
    \includegraphics[width=0.9\linewidth]{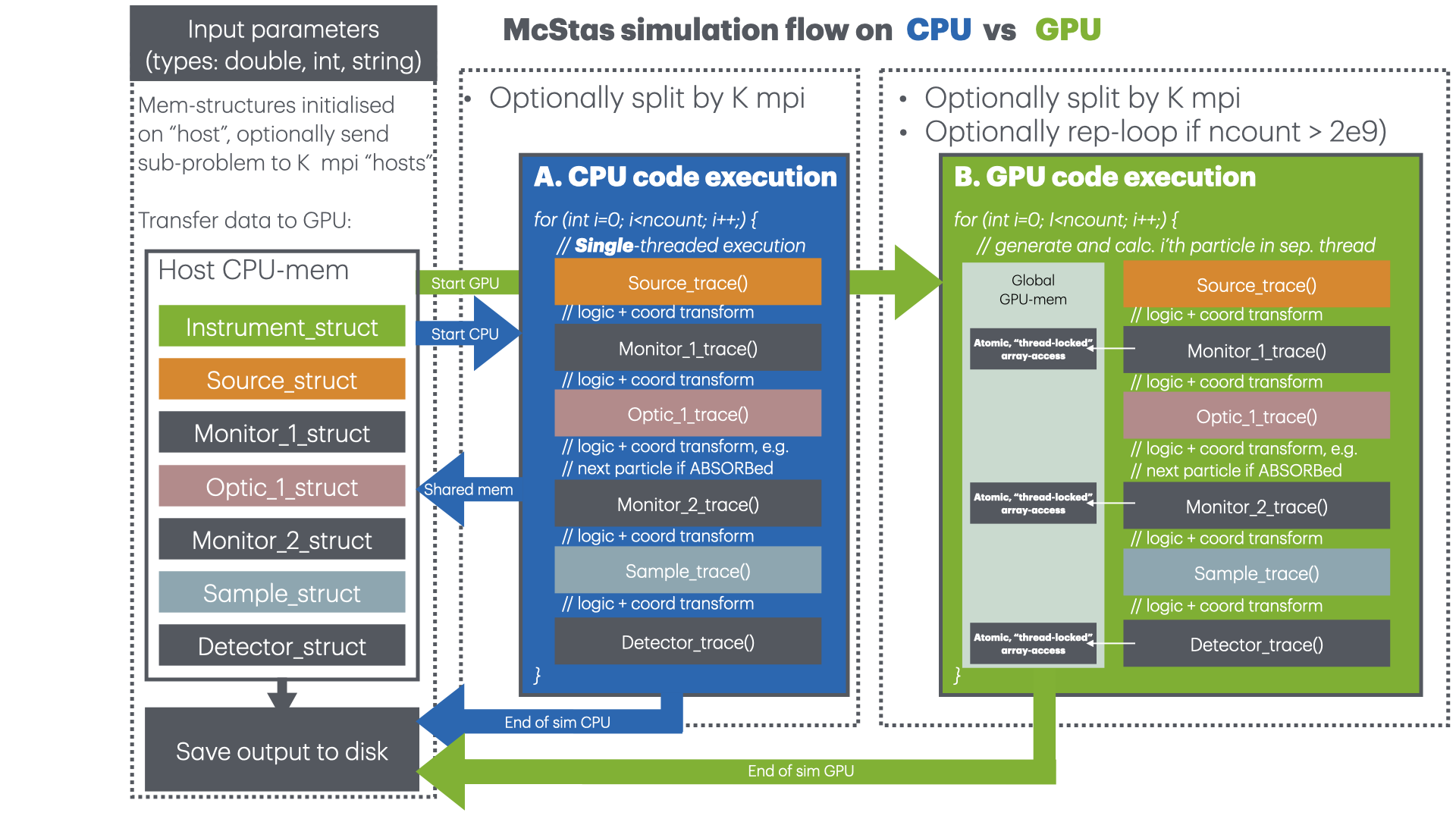}
    \caption{Schematic of a simulation flow in McStas, given a simulation statistic of \texttt{ncount} neutron rays. In the CPU execution setting (dark blue), the neutron-rays are treated one at a time in a single thread. In the GPU setting (bright green) the neutron-rays are executed in many parallel threads simultaneously, with thread-locked write-access to global GPU memory. Both the CPU and GPU simulation flow may be \textit{further} parallelized by means of MPI execution in a scatter-gather manner. }
    \label{fig:gpu-fig}
\end{figure}

Figure \ref{fig:gpu-fig} illustrates the McStas simulation flow during execution in three settings: single-threaded CPU, multi-threaded MPI, and parallel GPU execution. Given that McStas neutron-rays are treated in a fully independent manner, a simulation problem of \verb+ncount+ neutron rays is in principle of \textit{embarrassingly parallel} nature.
In the single-core simulation case, a single neutron is simulated at a time in a single CPU thread, with direct memory access to the host memory. In case of MPI on $N$ cores we apply a \textit{scatter-gather} approach where each of the $N_{cores}$ cores treat $\frac{\texttt ncount}{N_{cores}}$ of the original problem, with a following merge and save of the collected data on the MPI master node.
In the GPU implementation we instead run in a massively parallel execution where many neutron-rays are treated simultaneously, each in their thread. To ensure numerical validity and agreement with the CPU implementation, thread-locked access to the \textit{global} GPU memory was required and implemented via the OpenACC \verb+atomic+ clause. Transfers to and from the GPU happens at initialization time and at simulation finalization, as the host CPU handles input/output of files (Note also that a multi-GPU implementation was trivial: we simply scatter/gather the original problem like in the standard MPI case, sending $\frac{\texttt ncount}{N_{cores}}$ to each of $N$ GPUs).

To compare the acceleration capabilities, three different simulation schemes were tested:

\begin{enumerate}
    \item a single core simulation, to be used as reference;
    \item a Multi-threading approach with $N_{cores}=16$ cores using  MPI;
    \item a GPU approach with an NVIDIA A100.
\end{enumerate}

For each one of these schemes, all 70 models (listed in tables \ref{table::anisotropic_models} and \ref{table::isotropic_models}) were used for the comparison of the acceleration schemes. Each one was simulated in an independent virtual experiment with different number of incident neutrons ($N=10^5,10^7,10^8, 10^9, 10^{10}, 10^{11}$). The simulation run time was measured in every case.

\section{Results}

\subsection{Small angle scattering models}

\begin{figure}[h!]
    \centering
    \resizebox{1.0\textwidth}{!}
    {\includegraphics[]{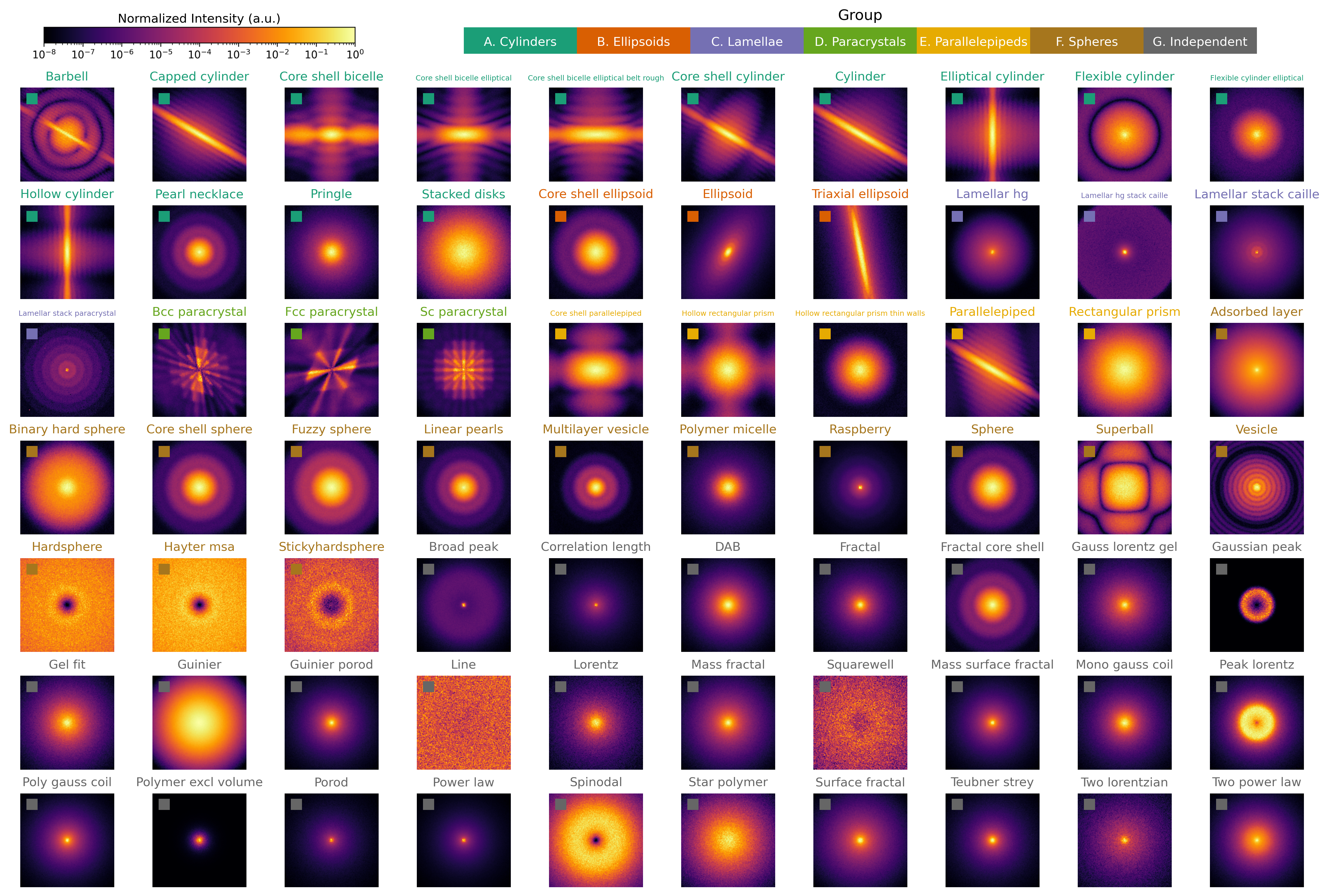}}
    \caption{Normalized scattering pattern results from SANS simulations in McStas for all models developed in this work. Each image corresponds to one simulation of the same instrument configuration but changing the sample. All samples were defined by setting the default parameters for each SAS model and no polydispersity. For all anisotropic models, no orientation distribution was included, resulting in perfectly oriented scattering objects to emphasize anisotropic scattering features. Rectangles indicate the groups in Tables \ref{table::anisotropic_models} and \ref{table::isotropic_models}.}
    \label{fig:models_sas}
\end{figure}

\begin{figure}[h!]
    \centering
    \resizebox{0.5\textwidth}{!}
    {\includegraphics[]{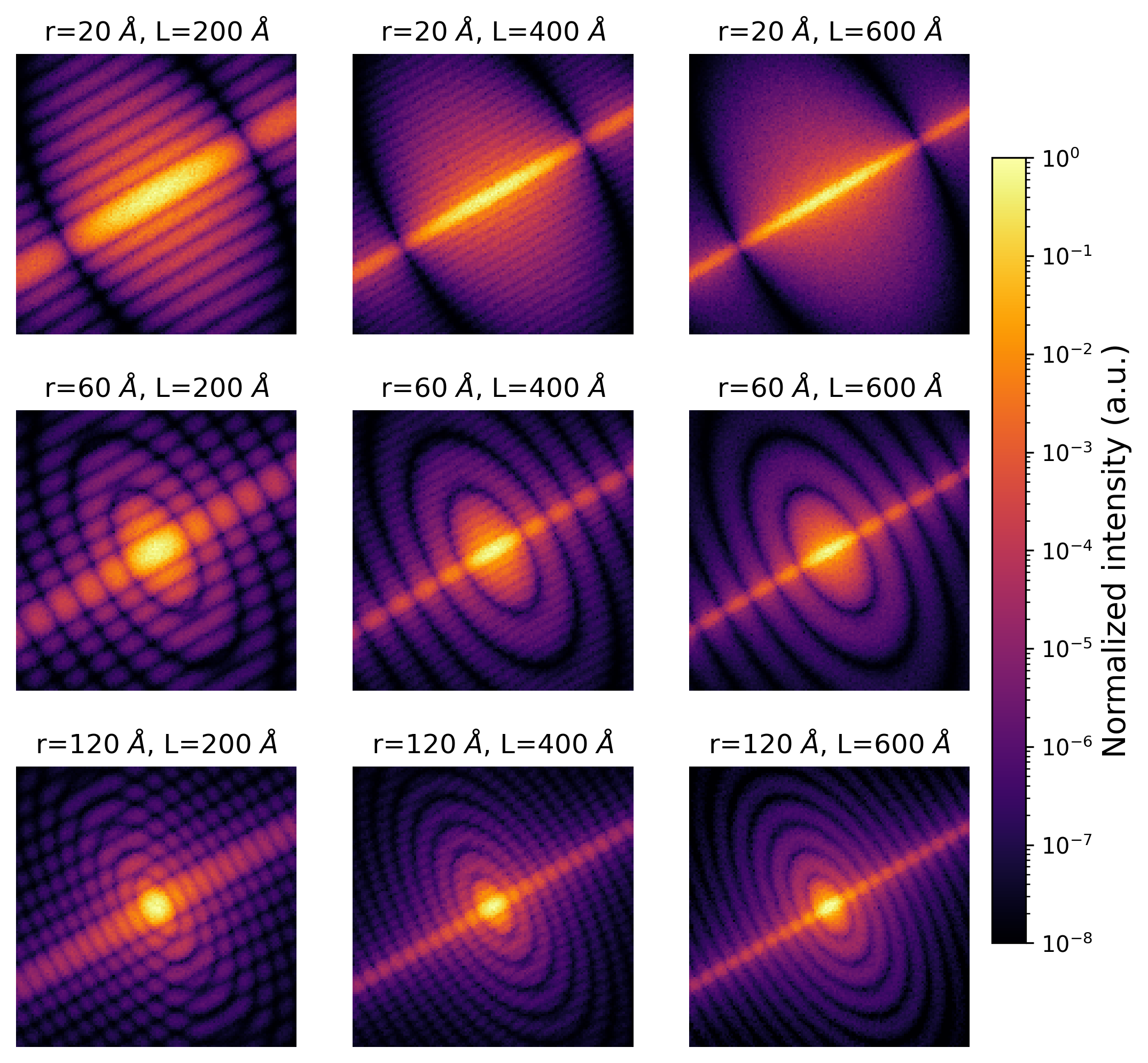}}
    \caption{Example of intra-class variability for a Cylinder form factor model with anisotropy. Each image corresponds to one simulation where the shape of the scattering objects is defined by the radius $r$ and length $L$ of the cylinders written above. In all simulations the cylinders were perfectly oriented at $\theta=\phi=60^{\circ}$ ($\Delta \theta = \Delta \phi = 0$) and were completely mono-disperse ($\Delta r = \Delta L = 0$).}
    \label{fig:intra-class}
\end{figure}

\begin{figure}[h!]
    \centering
    \resizebox{0.5\textwidth}{!}
    {\includegraphics[]{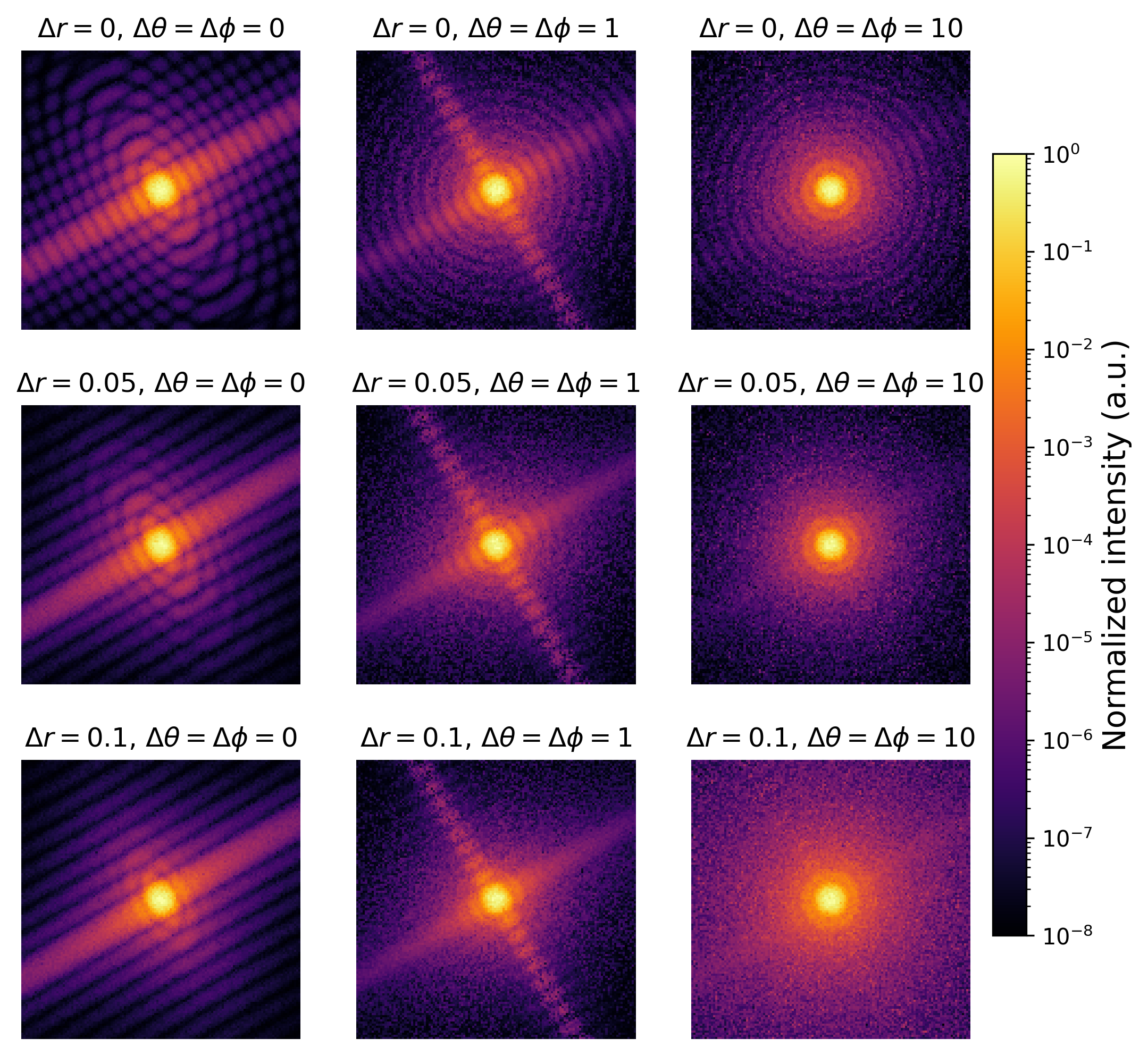}}
    \caption{Example of intra-class variability due to polydispersity in radius and orientational distributions for a Cylinder form factor model with anisotropy. Every image corresponds to a different simulation with varying parameters as described above the image and common parameters mean radius $r=120\,\AA$ and cylinder length $L=200\,\AA$.}
    \label{fig:poli}
\end{figure}

The results of 70 simulations in which the instrument configuration was fixed and the sample description was changed in all of the models presented in this work are shown Figure \ref{fig:models_sas}. Each scattering pattern corresponds to testing one of the SAS models $M_m$ listed in Tables \ref{table::isotropic_models} and \ref{table::anisotropic_models} with the default values for the model parameter definition of the component in McStas. These default values are directly obtained from SasView, and have been selected since they are good prior estimate for reasonable physical models (they fulfill constraints and are estimated towards user needs in real scenarios).
The scattering patterns in Fig. \ref{fig:models_sas} show the variability in neutron intensity that can be present at the detector when selecting the SAS model $M_m$. This variability can be referred to as inter-class variability. Some models are clearly different than others, and some others look very similar with the parameter definition used to generate Fig. \ref{fig:models_sas}. Nevertheless, intra-class variability can also provide very different scattering patterns for the same model $M_m$. To exemplify this, the parameter space of the anisotropic model \emph{Cylinder} is varied drastically on different simulations and the resulting scattering patterns are shown in Figs. \ref{fig:intra-class} and \ref{fig:poli}. In this example, the model considers no structure factor (dilute regime) and only the form factor contributes to the anisotropic scattering pattern. To better understand these figures, the analytical expression of the cylinder form factor is revised. Following Eq. \ref{eq:form_factor}, the form factor of a cylinder can be obtained by defining the scattering amplitude as \cite{guinier, nist}

\begin{equation}
A_{cylinder}(Q_{AB},{Q_C}|\rho, \rho_s, r, L) = 2 (\Delta \rho) V \frac{\sin{\left(\frac{1}{2}Q_{AB}L\right)}}{\frac{1}{2}Q_{AB}L}    \frac{J_1(Q_{C}r)}{Q_{C}r},
\end{equation}
where $J_1(x)$ is the first order  Bessel function, $\Delta \rho$ is the excess of SLD of the cylinder and the solvent, and $r$ and $L$ are the cylinder's radius and length respectively.  The volume is a function of the parameters and must be calculated, in this case for a cylinder $V_{cylinder}=\pi r^2 L$. In the case of the McStas simulation, the parameter space that can be varied is larger than that of the form factor definition given since the polydispersity and orientational distributions must be defined as explained in section \ref{sec:poli}. In this case the model parameter space is $\vec{\theta}_{cylinder}=\{\rho, \rho_s, r, L, \theta, \phi, \Delta r, \Delta L, \Delta{\theta}, \Delta{\phi}\}$, where $\rho$ and $\rho_s$ are the SLD of the cylinder and solvent respectively, $\theta$ and $\phi$ define the orientation angles with respect to the incident neutrons (and therefore the projections $Q_{AB}$ and $Q_C$), and $\Delta x$ is the corresponding relative variance parameter of parameter $x$ that defines the variance of the  polydispersity or orientational distribution. To simplify the interpretation of each parameter, the results presented in Fig. \ref{fig:intra-class} are obtained for fully oriented mono-disperse cylinders ($\Delta x=0$ for $x=r,L,\theta,\phi$) and Fig. \ref{fig:poli} shows the effect of polydispersity and orientational distribution for fixed form factor model parameters. 

As can be seen in both, Figs. \ref{fig:intra-class} and \ref{fig:poli}, the scattering patterns corresponding to a given model may be very different given that all parameters express themselves in different features on the 2D scattering pattern. The variation of model parameters $r$ and $L$ shown in Fig. \ref{fig:intra-class} exemplifies how the shape of the scattering objects can produce varying scattering patterns even though they arise from fully oriented cylinders. Adding variation on the orientation of the cylinders transforms an anisotropic scattering pattern towards an isotropic one (first row on Fig. \ref{fig:poli}), while adding polydispersity to a shape parameter blurrs the characteristic oscillations corresponding to it (evanescent ellipses on the first column in Fig. \ref{fig:poli}). Given that in these simulations all the cylinders were of the same length $L$, the spacing of the diagonal line features resulting from oriented particles of a fixed length $L$ do not vary unless the relative orientation between cylinders is lost.

\begin{figure}[h!]
    \centering
    \resizebox{0.8\textwidth}{!}
    {\includegraphics[]{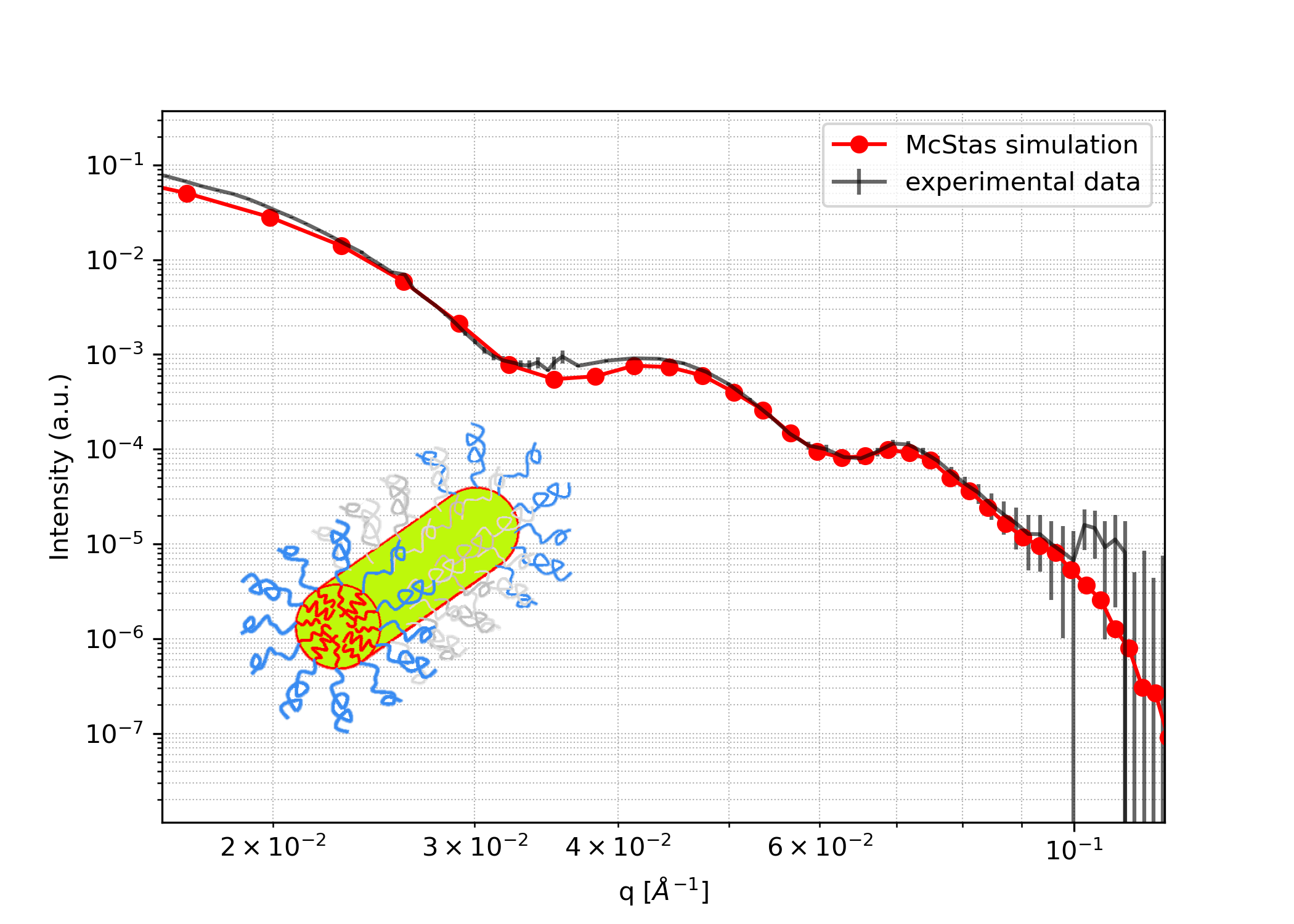}}
    \caption{Comparison of small angle scattering curves of a cylindrical model McStas simulation with experimental data  measured at KWS-1, FRM-II. Inlay image depicts the amphiphilic polymers POO$_{10}$-PEO$_{10}$ forming the scattering object, which by contrast matching can be considered of cylindrical shape.}
    \label{fig:benchmark}
\end{figure}

Finally, we benchmark the cylindrical model with an experimental data curve of cylindrically shaped micelles measured at the SANS instrument KWS-1 located at FRM-II \cite{feoktystov2015kws}. This dataset has been widely studied since it is part of the Lab-course material of the J\"ulich Centre for Neutron Science \cite{labcourse}. The sample consist on amphiphilic  polymers POO$_{10}$-PEO$_{10}$ in heavy water as shown in the inset of Fig. \ref{fig:benchmark}. During the experiment, the contrast was matched so that only the core of the micelle was visible for neutrons. The measured data is plotted with error-bars in  Fig. \ref{fig:benchmark} (black curve). 
To test our cylindrical form factor model, the simulation of the McStas instrument of KWS-1 published in \cite{robledo2024} was used to describe the instrument, and the novel component of the cylindrical model (A.7 in Table \ref{table::anisotropic_models}) was selected to describe the sample. Given that the position of the lowest Q drop in the experimental data is $Q_{min}=0.032\AA^{-1}$, and using the relation for cylindrical form factors of $R=3.78/Q_{min}$ \cite{hammouda2008sans}, then the radius of the cylinder sample was set in the McStas simulation at 119 $\AA$. For the cylinder length, the experimental data was fit using SasView and constraining the radius to the previously obtained value. From this fitting a value of $L=800\,\AA$, and polydispersities of 5\% for the radius and length parameters were used in the Monte Carlo simulation. In addition, the orientation of the cylinders was set to random. The result of this McStas simulation can be seen in Fig. \ref{fig:benchmark} with red circles. The agreement between the curves is an indicator that it is now possible to simulate a large variety of types of sample by means of the McStas SAS models presented in this work.\\

\subsection{Acceleration schemes}

\begin{figure}[h!]
    \centering
    \resizebox{0.8\textwidth}{!}
    {\includegraphics[width=10cm]{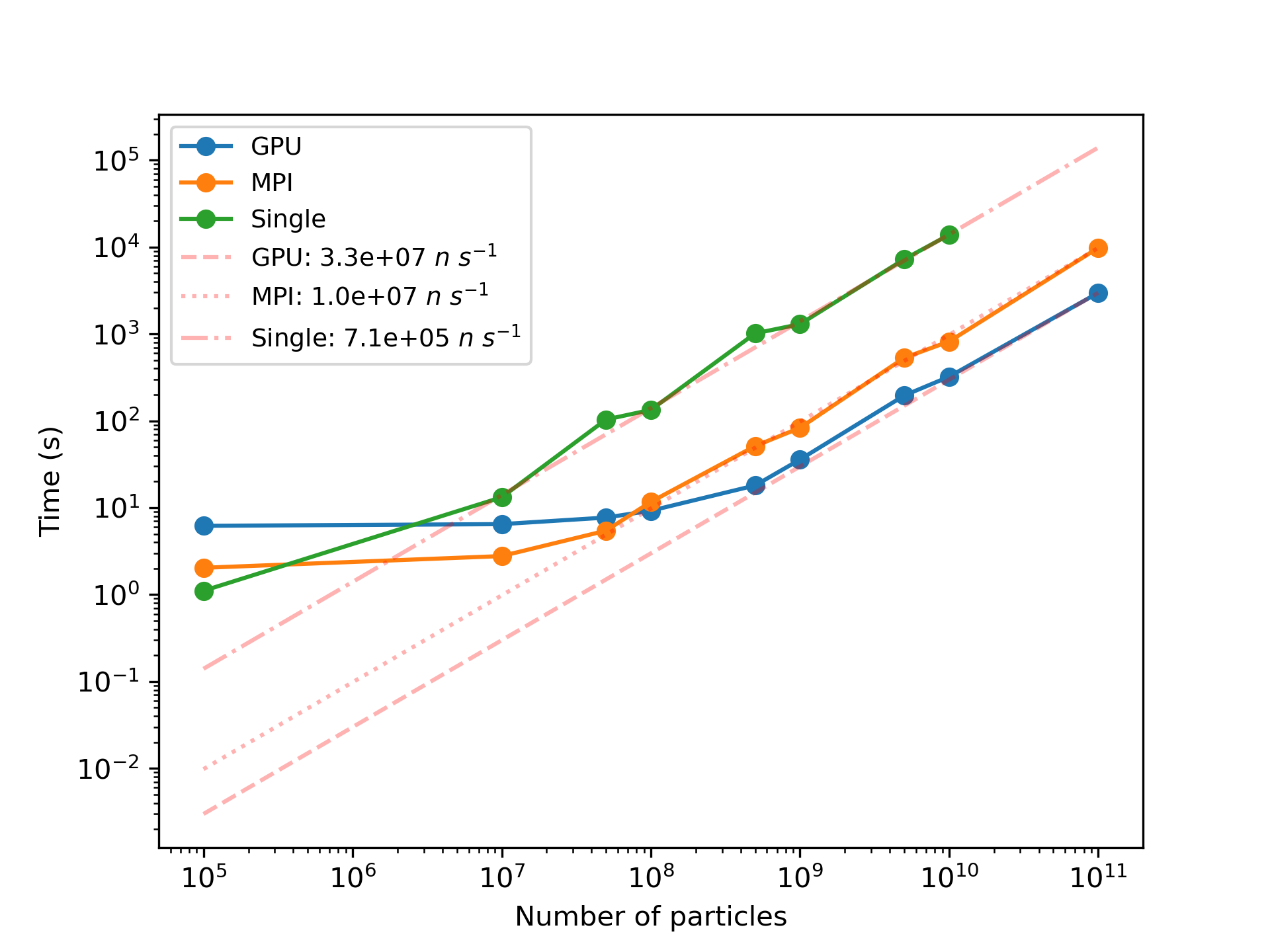}}
    \caption{Median time of the models presented in this work as a function of the number of simulated neutrons, using the Graphical Processing Unit (GPU), Message Passing Interface (mpi) or single core (single). The inverse of the slope, in neutrons per second, is given in the labels for each type of simulation.}
    \label{fig:median_time}
\end{figure}

All models described in Tables \ref{table::anisotropic_models} and \ref{table::isotropic_models} were tested in simulations with different numbers of incident neutrons $N$ under the three acceleration schemes described in section \ref{sec:acc}. Their corresponding run times were measured. Since a motivation for accelerating MC simulations is the ability to improve statistics by scaling the number of simulated neutrons $N$ to improve MC estimates, the median value of the run time of all models as a function of $N$ is shown in Fig. \ref{fig:median_time}.
At low number of particles $N$, the cost in time of threading may be higher than the gain in computation time of the actual parallelized run, resulting in a faster single core simulation. In the case of the GPU approach, the tasks of uploading and downloading data is very time consuming, resulting in the worst performance. Therefore, for small number of particles ($N<10^6$), choosing a single core simulation is the best choice. At intermediate $N$,  both the GPU and MPI approach start to show advantages while they get filled with  neutron trajectories, and the cost of distributing the tasks becomes smaller than the time gained by parallelization. We can conclude from Fig. \ref{fig:median_time} that an MPI acceleration scheme is the optimal choice in the range $N \in \left[10^6, 10^8\right]$. For larger values of $N$ ($N>10^8$), optimizing simulations by including GPU parallelization gives the best performance among the three methods discussed, in this case being approximately 3 times faster than the 16 core MPI approach. Scaling to larger amounts of cores or GPUs in a cluster will lead to increasing accelerations and the run-times of these models can be estimated by means of the results presented in this work. The acceleration schemes presented here are those that can be found in a personal computer.

When either the GPU or all of the cores available get filled with trajectories, then the simulation time increase becomes linear with increasing $N$. The slope values of the linear fits presented in this figure are shown in the labels of Fig. \ref{fig:median_time}. Compared to a single core simulation, the MPI approach with 16 cores results in a speed up ratio ($t_{single}/t_{mpi}$) of $14.26$. This can be interpreted as if each core is resulting in a net factor of 0.89 of the calculation power of a single core. The deviation from a 16 times gain using 16 cores is given by the fact that the threading and summation of results are time consuming. On the other hand, by using one single NVIDIA A100 (40GB) GPU, the speed up ratio is of 46.87, which can be thought as equivalent to using 52 cores in an MPI simulation. As mentioned before, the deviation from the linear tendency at low number of simulated particles is due to the GPU having capacity for more trajectories.

It is interesting to observe that the median times of the distributions cross somewhere in between $5\times10^7$ and $5\times10^8$ neutrons. Also, the increase from $1\times10^7$ neutrons to $5\times10^8$ is almost costless in computation time for all the models when using the GPU. This is a consequence of the under-use of the full capacity of the GPU for low $N$. The asymptotic value at low $N$ of about 5 seconds for the mean (among all models) runtime is an estimator of the time needed by the GPU card to upload and download the simulation data by means of OpenACC. 

\begin{figure}[h!]
    \centering
    \resizebox{0.8\textwidth}{!}
    {\includegraphics[]{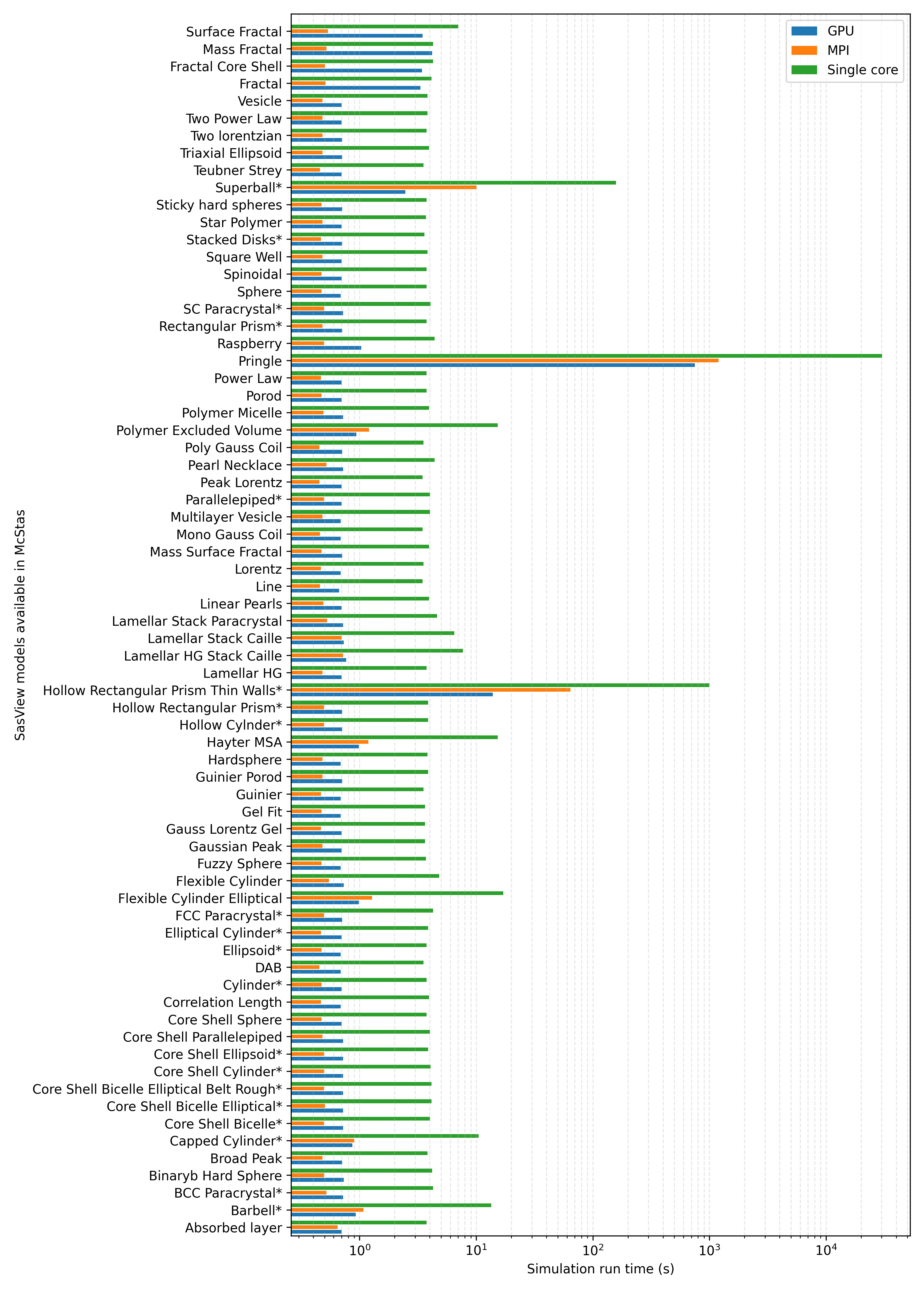}}
    \caption{Simulation time of $N=1\times10^8$ neutrons in McStas for each of the  70 analytical SANS models described in SasView (now available in McStas 3.4). Colors indicate the different speed-up methods employed: GPU stands for a simulation using the Graphical Processing Unit (GPU) NVIDIA A100 (40GB) by means of OpenACC, MPI for multi-threading using OpenMPI, and Single core for a sequential simulations using only one thread in a single CPU. The asterisk in name labels indicate the anisotropic models.}
    \label{fig:time_per_model}
\end{figure}

Fig. \ref{fig:median_time} also shows that simulations accelerated by GPU (blue curve) take roughly the same time as those accelerated by MPI (orange curve) for $N=1\times10^8$. We decided to study the simulation run time variation as a function of the model component and the acceleration scheme for this particular value of $N$. The results, presented in Fig. \ref{fig:time_per_model},  show that both optimization schemes presented are more convenient than the single core sequential simulation. It can be seen that the multi-threading approach is more convenient than the GPU approach in the majority of the SAS models. However, in those computationally expensive models, such as Barbell, Capped cylinder, Hollow Rectangular Prism Thin Walls, Flexible Cylinder Elliptical, Hayter MSA, and Polymer Micelle, the GPU calculation with OpenACC is faster than the multi-threading one with MPI. For higher values of $N$, all models run faster on GPU than on MPI. This figure also provides a relative estimate of the complexity, and therefore on the execution time, of the components presented in this work and can be used as a reference when selecting the adequate model for a simulation.  

\section{Conclusion}

This work describes the optimized SAS sample components that are readily available in McStas, which opens up the landscape for the possible small angle neutron scattering virtual experiments that can be performed with the software. All the SAS components have the effects of polydispersity as well as orientational distribution included. Components can be used under acceleration schemes, according to the necessity. For single virtual experiments, where statistics is not very important, rather qualitative features, single core simulations are adequate. When increasing the number of particles in a given simulation, multi-core and GPU parallelization of the particle traces becomes relevant. The GPU parallelization outperforms the multi-core simulation presented in this work when the number of neutrons in the simulation is higher than $10^8$. This can be useful if attempting to observe low probability scattering effects on simulations where neutrons perform very long trajectories. Of course, by increasing the number of threads of MPI with enough cores, or the number of GPUs used, the speed-up gains of each method can be exploited even stronger. But in this work, a reasonable value of threads as well as number of GPUS that a current notebook can offer was chosen for comparison.

The parallelization of traces in Monte Carlo particle tracing algorithms can be deeply exploited in simulations with large hyper-parameter spaces to explore a vast region of this space and generate large datasets of Monte Carlo simulations. Simulation datasets can help in data augmentation \cite{gans_xs},  as well as to learn from them through machine learning and then to infer on small datasets of real experiments (which is the usual case in neutron scattering experiments) \cite{FRANKE20182485}. With the recent advances in Generative Adversarial Networks that allow to map from a simulated data distribution to an experimental data distribution in a bijective manner \cite{gans_ns}, this type of mechanism for dataset generation can become very useful.

\begin{acknowledgments}
This work has received funding from the European Union's Horizon 2020 research and innovation programme under the Marie Skłodowska-Curie grant agreement No 101034266.

This work benefited from the use of the SasView application, originally developed under NSF award DMR-0520547. SasView contains code developed with funding from the European Union’s Horizon 2020 research and innovation programme under the SINE2020 project, grant agreement No 654000.
\end{acknowledgments}

We would like to acknowledge the Helmholtz Artificial Intelligence group by providing computational resources through HAICORE project 27114.

\bibliographystyle{unsrt}
%\bibliography{bibliography}

\end{document}